**Influence of the pH on the synthesis of carbon quantum dots and graphene quantum dots with applications potential nanosensors**

Pedro Holland de Oliveira, and Ana Champi[*]

Laboratory of New Carbon Materials: Graphene, Center of Natural and Human Sciences, Federal University of ABC, Santo André, SP 09210-580, Brazil

[*]Corresponding author: ana.champi@ufabc.edu.br

## Abstract

In this work, we confirmed the presence of carbon quantum dots (CQDs) by optical techniques such as UV-visible and photoluminescence (PL), after filtering the result of carbonization and sulfonation of microcrystalline cellulose (MCC), which are synthesis processes for obtaining these nanomaterials. The high-resolution transmission electron microscopy (HR-TEM) confirmed, not only the size of the particles, but the crystalline pattern of the carbon-based nanoparticles. The HR-TEM also indicates the presence of graphene quantum dots (GQDs), since the presence of anisotropic shaped particles was observed. Fluorescence and absorbance indicate the pH dependence on optical and structural properties, as well as on the production of CQDs and GQDs. The pH may be influencing the syntheses of CQDs and GQDs and the origin of fluorescence in these systems, due to presence of the functional groups such as hydroxyl, carboxyl, carbonyl, and sulfoxide at low pH, and in the $sp^2$ regions of the edges. Finally, CQDs and GQDs may have potential applications as nanosensors due to the easy functionalization of their functional groups with biological systems, which can be detected by shifts in photoluminescence spectra, demonstrating potential for applications as nanosensors.

## 1. INTRODUCTION

Carbon Dots (CDs) are a group of nanoparticles composed mostly of carbon atoms. These materials present interesting properties, related to their size and functional groups present in their surface, such properties, as conductivity and fluorescence, enable their application in a bunch of different areas, for example, Electrochemical Capacitors, enhancing capacitive performance [1, 2], and applications in biomedicine, light emitting diodes (LEDs), sensing and many others, using specific properties, as low toxicity, luminescent properties, and fluorescence quenching effect [3, 4]. For that to be possible, it is important to understand what causes these properties, and how to increase them, as an example, increase conductivity in carbon nanodots [2].

Some nanoparticles, as the CDs, are considered to be zero dimensional (0D) [5, 6]. This notation refers to the dimensions of the material that are confined, which represents a quantization on the energy levels, in each dimension. For quantum dots this effect occurs in all three dimensions. With the discretization of the energy levels, the material present changes in electronic and optical properties [5]. This explains the size dependence of size of their properties, for example, the change in emission fluorescence wavelength depending on the size of the nanoparticle. In general, carbon dots refers to any carbon-based material with less than 10 nm [3].

Carbon Dots (CDs) include a variety of different nanomaterials, all zero dimensional and carbon-based. It is variations may change in size, chemical functionality and other characteristics [3]. To distinguish these possible variations, some classes of CDs were created, consisting in three main categories: The first one is the Graphene

Quantum Dots (GQDs), consisting in one or few layers of graphene, and do not present radial symmetry. The second group, and the one of interest in this work, are the Carbon Nanodots (CNDs), which may present some degree of crystallization and ordered layers, if they are spherical shaped, and have a clear crystal lattice, they are called Carbon Quantum Dots (CQDs) [3].

In particular, CQDs and GQDs are very promising materials for biosensors applications, an application of interest is as fluorescent nanosensor. This type of sensor detects the presence, and quantifies the amount of a biological systems through the quenching of it is fluorescence. Different nanoparticles are already in use for such detections, including quantum dots [7, 8], and CQDs have also been used is nanosensors [9]. The application of this materials for this type of sensor rely on the presence of functional groups in the surface (as hydroxyl, carboxyl, carbonyl and sulfoxide) of these nanoparticles, these groups usually contribute to the fluorescence, and the interaction with the biological systems, causes a quenching effect [9, 10].

On the other hand, for the synthesis of these materials, two possible approaches are used: The “top-down” methods consist on fragmenting carbon-based precursors, to create these nanoparticles, which can be done physically or chemically, while the “bottom-up” methods are the opposite, consisting in the combination by polymerization or carbonization of molecules generating the CDs [4, 11, 12].

In this work, a "top-down" strategy is used to obtain CQDs, and a study on the influence of pH was carried out on these systems. Due to the use of sulfuric acid during the synthesis process, the resulting nanomaterial is highly acidic, which promotes the formation of various functional groups and can also influence the exfoliation of graphitic monolayers during the washing process with ultrapure water to increase the pH. The latter can occur similarly to the obtaining of graphene oxides from the modified Hummer method, in which chemical exfoliation is performed, including to obtain graphene chemically [13,14]. In this direction, we work with two pHs to obtain the CQDs (pH = 0) and pH = 4 to obtain GQDs. The nanomaterials were characterized by UV-Vis, photoluminescence, and HR-TEM to study their optical, electronic and structural properties. Finally, HR-TEM confirm the exists of QCDs and GQDs which is more promisor to potential applications like nanosensors.

## 2. EXPERIMENTAL PROCEDURES

### 2.1 CQDs

#### 2.1.1 Materials

Microcrystalline cellulose (MCC, 100% cellulose) from Sigma-Aldrich. Sulfuric acid ($H_2SO_4$, 98%) was acquired from LabSynth to synthetize the CQDs.

Ultrapure (type 1) water from the Milli-Q Direct-Q 3UV was used for the preparation and washing of all nanostructures. Water pH was measured as around 5,6.

### 2.2 CQDs Characterizations

The CQDs samples were characterized by UV-Visible, Photoluminescence (PL) and HR-TEM.

The UV-Visible spectroscopy (absorbance) was used to understand the optical properties of the CQDs, the equipment used was a Thermo Scientific Evolution One Plus spectrophotometer, in the range 190-800 nm.

The emission spectra (PL) was measured to understand the fluorescence of the CQDs, which can also provide information of their structures, the equipment used was a Varian Cary Eclypse fluorescence spectrophotometer, in ranges of 350-600 to emission, and 300-450 to excitation.

At last the high-resolution transmission electron microscopy (HR-TEM) of the sample with pH 4, allow a deeper understand of the physical structure of the CQDs, as well as their size and shape, the equipment used was a Talos F200X G2 Transmission Electron Microscope.

### 2.3 Synthesis of CQDs

These carbon dots are of the carbon quantum dots category, they were synthesized using a simultaneous carbonization and sulfonation of the microcrystalline cellulose [15]. This method is a "top-down" way of obtaining the CQDs. The reaction of the MCC and the $H_2SO_4$, will result in the creation of an solid acid catalyst (SAC), The carbon dots are then obtained by rinsing this solid, resulting in the detach of these nanoparticles [12].

First, 2 g of MCC was added to 40 g of $H_2SO_4$, while under magnetic stirring until the solution was homogeneous. This solution was then heated to 150 °C in an oil bath for 1 hour. After that this mixture had cooled to the ambient temperature, water was added and the new mixture was centrifuged at 3.000 RPM. The supernatant was then collected and filtered in a syringe filter of 0,22 µm, the filtered solution obtained is the first sample (with pH = 0). This rinsing procedure (addition of water, centrifuge, and filtering) was repeated in the solid residue until a pH 4 was obtained, for that 21 rinsing were needed, the last supernatant, after filtering is the second sample (pH = 4).

Both samples were observed under a UV light, and fluorescence was observed, as seen in Figure 1. The samples were named CQDs (pH = 0), for the carbon dots at pH = 0, and CQDs (pH = 4), for the carbon dots at pH = 4, these names were used for knowing that the particles are carbon quantum dots, since the characterizations showed their crystalline pattern.

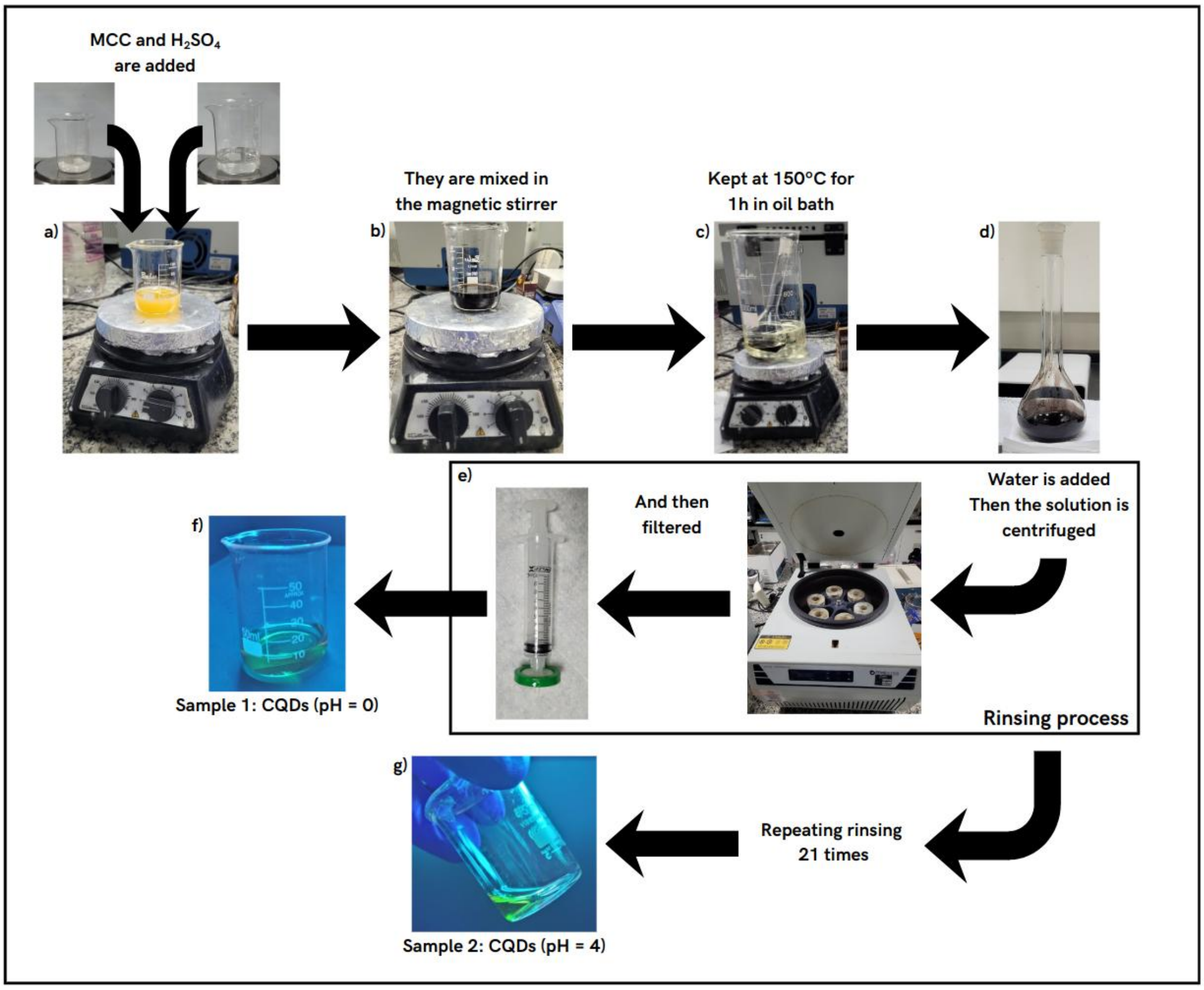

Figure 1: Synthesis of the two samples of CQDs. a) The MCC is slowly added to the $H_2SO_4$. b) They are mixed until the solution is homogeneous. c) Heated to 150ºC for 1h in oil bath. d) The solution is let to cool down. e) Water is added, then the solution is centrifuged, and the supernatant is filtered (Rinsing process). f) The first sample is obtained by rinsing the solid fraction once: CQDs (pH = 0). g) The second, by rinsing it 21times in total: CQDs (pH = 4).

## 3. RESULTS AND DISCUSSION

### 3.1 UV-Visible

The first indicative of the presence of carbon dots in the liquid fraction was the fluorescence under UV light, as observed in Figure 1f, g.

To understand the optical properties of the CQDs, the three absorption bands are analyzed. The band around 200 nm is associated with π-π* transitions in the $sp^2$ aromatic domain, this transition occurs at the core of the CQDs [16, 17]. It is possible to see that this first band present the most intense absorbance for both samples, and a shift towards blue with increasing pH.

Another band can be seen in both samples, at around 285 nm (Figure 2a), this band is associated with n-π* electronic transitions, this are edge transitions of the CQDs, this transition connects the functional groups as hydroxyl (–OH), carboxyl (–COOH), carbonyl (–CO) and sulfoxide (–SO) due to synthesis process [10], at the surface of the CQDs to the orbitals in the $sp^2$ domain. In this case, it is possible to see that this second band is more intense in the CQDs (pH = 0) sample, this is related to the variations in the third band, and the alterations in the photoluminescence.

The last band seen, around 360 nm, is also associated with n-π* transitions, in this case, the band is associated with surface states, related to the functional groups and doping heteroatoms. This band can be seen more clearly in the CQDs (pH = 4) sample, but it is present in both samples.

In the Figure 2b we can see the Tauc plot for the same CQDs samples, it was possible to see that the band gap of these nanostructures is direct. From the linear extrapolation to zero [18, 19], the optical band gaps of these structures are measured, resulting in 2,9 eV for the CQDs at pH 0, and a reduction for the increasing pH, at pH 4, the band gap measured is 2,7 (this difference corroborate with the wavelengths observed in the PL measures, as seen in Figure 4).

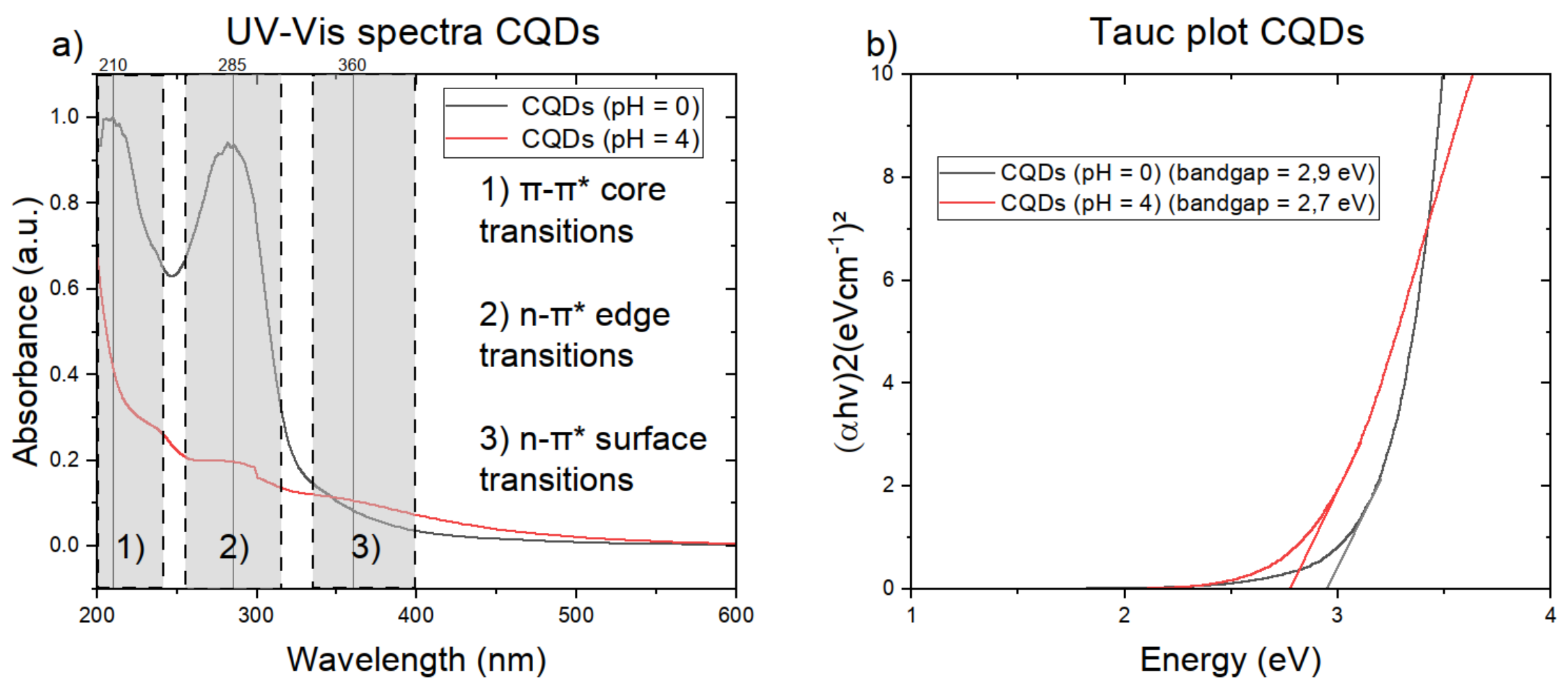

Figure 2: a) UV-Visible spectra of both samples of CQDs. b) Tauc plot of both samples of CQDs.

### 3.2 Photoluminescence (PL)

In the PL map of both samples (Figure 3) it is possible to see the regions of peak excitation and emission. Three major differences can be seen at the between them, the first one is the intensity of the fluorescence response, the second is a broad spectrum in the rinsed sample (pH = 4), and the third is a redshift in the emission region when the pH increases.

This differences alongside the UV-Vis results shows that the Carbon Dots under acidic conditions present protonation on the edge sites, quenching the fluorescence observed from these regions [20]. This causes a limitation on the CQDs (pH = 0) sample, the fluorescent response related to the functional groups in the surface is responsible for most of the signal obtained, which is the cause for a broader spectrum in the CQDs (pH = 4) sample.

This also explains the difference in intensity in the UV-Vis bands, the CQDs (pH = 0) absorbance band around 360 nm is not visible, since it is the region of peak excitation for this sample. The sample CQDs (pH = 4) also presents its peak at this region, but with a lower intensity.

Also, for the 285 nm absorbance bands, the CQDs (pH = 0) do not present fluorescence response in this wavelength and, therefore, it is a very intense band. As for the CQDs (pH = 4) sample, this band is not as intense, and, as seen in Figure 3b, this samples excitation is still intense in this region.

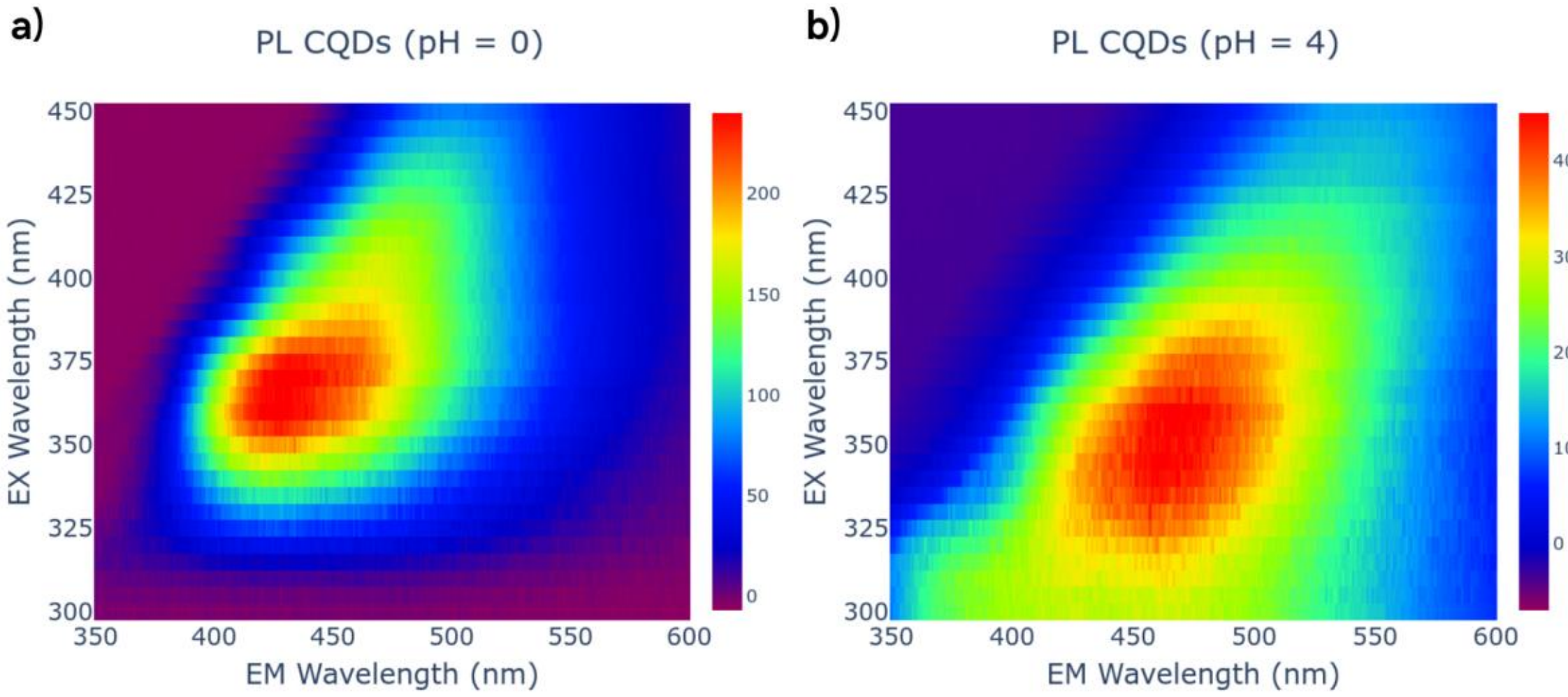

Figure 3: Emission map of photoluminescence (PL) spectra of both samples of CQDs. a) Sample pH = 0. b) Sample pH = 4.

As seen in Figure 4, the excitation associated with peak emission for the pH = 0 sample is 370 nm, and for the pH = 4 is 360 nm, while the emission peak presents a redshift with increasing pH, which is in accord the variations observed in other works [15].

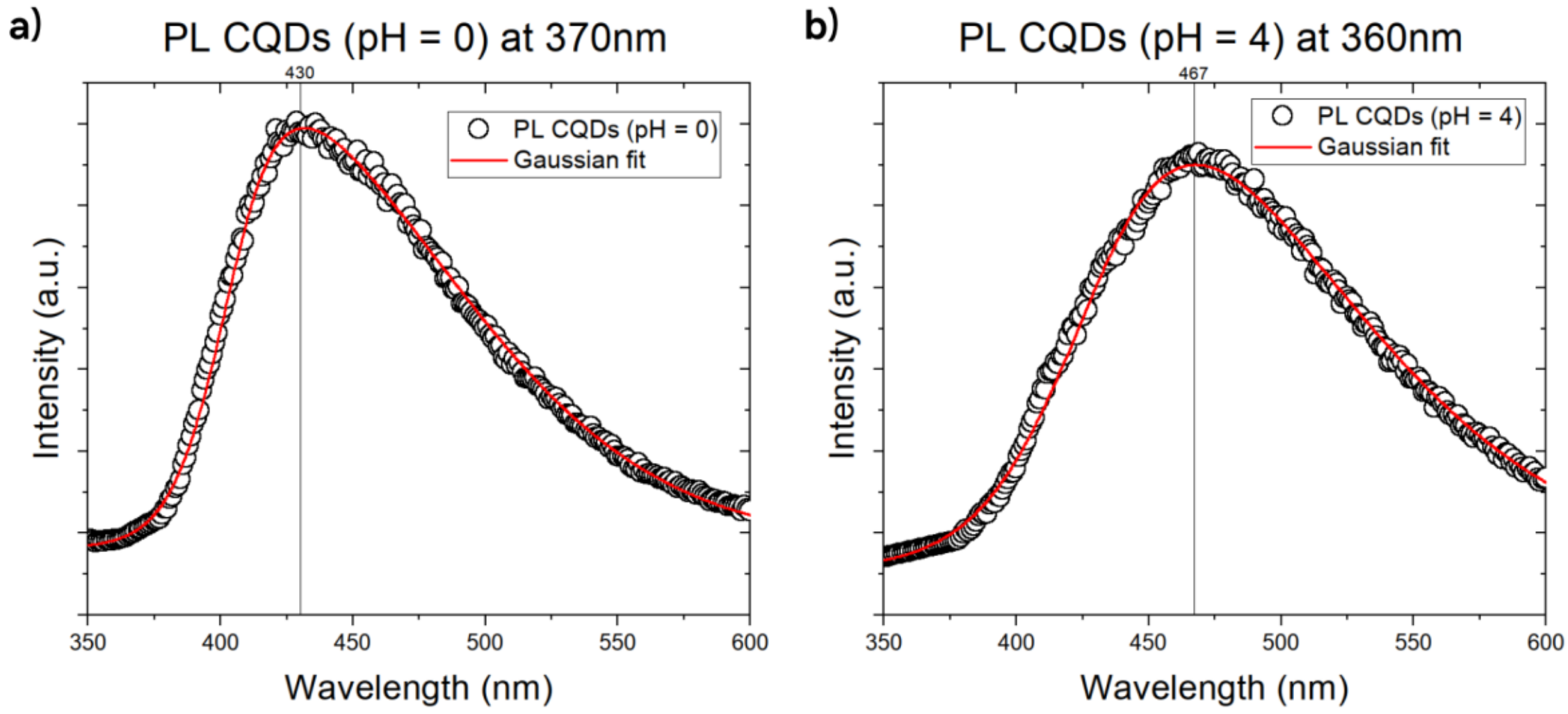

Figure 4: Maximum emission band of photoluminescence (PL) spectra of both samples of CQDs. a) Sample pH = 0. b) Sample pH = 4.

Since the peaks of fluorescence, for both samples, rely on the functional groups, this again brings the possibility of an application as a nanosensor. The increase in pH, needed for the interaction with most biological systems, enable an increase in fluorescence of a broader spectrum of wavelengths, even so, the main excitation still resides on the functional groups present, which means that a quenching effect would be simple to measure, facilitating an application like this [9].

### 3.3 HR-TEM

The high-resolution transmission electron microscopy allows an understand of the physical structure and morphology of the nanoparticles. In the Figure 5a, it is possible to see that these nanoparticles have less than 10 nm, which confirms the presence of carbon dots, and the crystalline pattern, since the layers are visible in this image, this suggests that the particles are carbon quantum dots (CQDs) [3].

Some interesting results are shown if Figure 5b, it is possible to see that some of the particles are not spherical, but have an anisotropic shape, in the two dimensions seen in the figure, one dimension is almost twice the size of the other (approximately 1.9 times the size). This characteristic may indicate the presence of, not only CQDs, but also graphene quantum dots (GQDs) [3]. This possibility creates a lot of new possibilities, since the GQDs have specific electrochemical properties, as ultra-high scan rate, and low specific capacitance, with possible applications in fuel cells and lithium ion batteries [21].

The crystalline lattice spacing of 0.2 nm, which is close to the graphene lattice spacing, in the plane (100) [3], confirming that the carbon dots obtained were CQDs, and potentially GQDs.

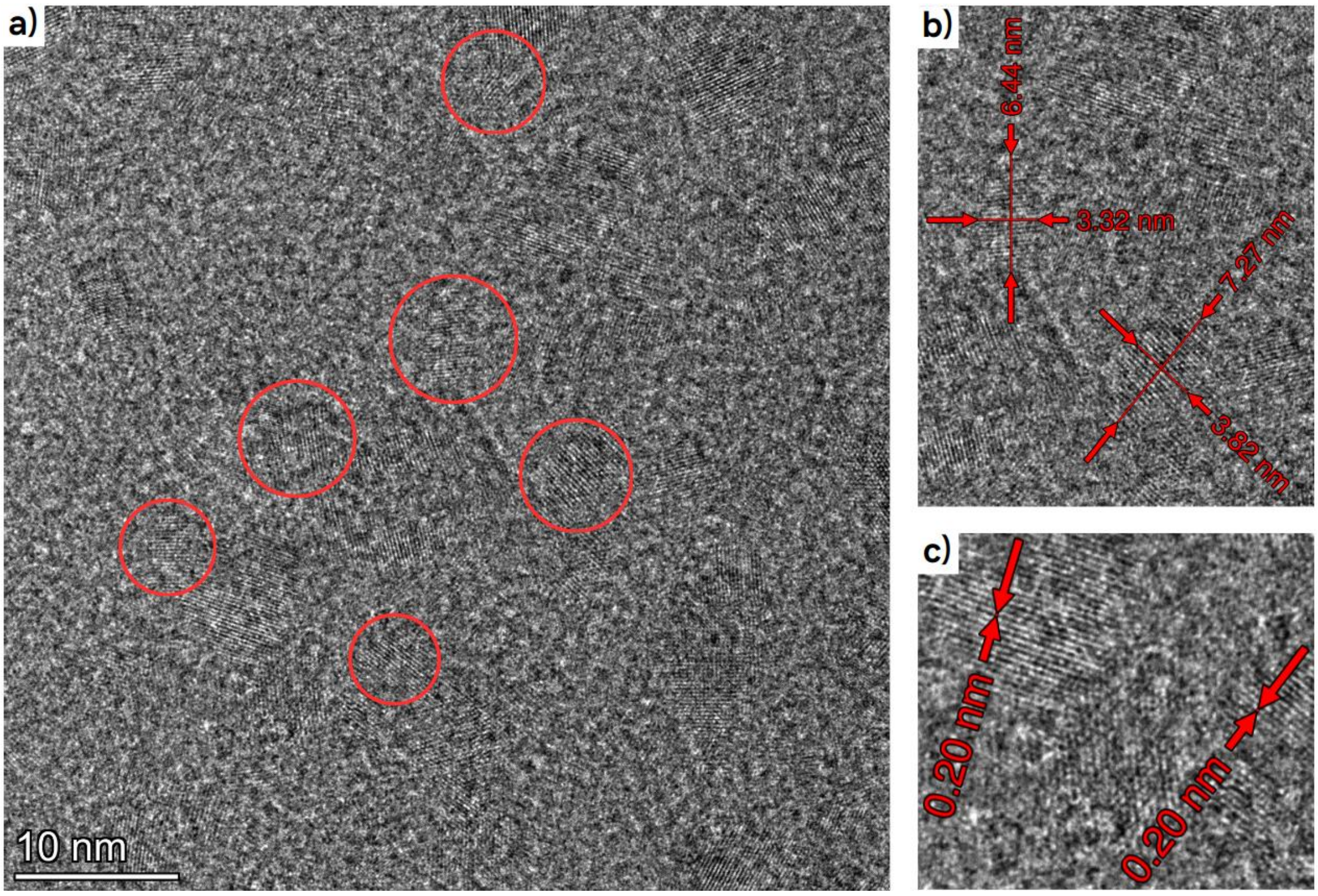

Figure 5: HR-TEM of CQDs (pH = 4). a) Particle distribution. b) Particle size. c) lattice spacing.

## 4. CONCLUSIONS

This study demonstrates that the synthesis processes used are successful in producing carbon dots, and confirming the crystalline structure of graphene layers, it corroborates that the carbon dots are indeed carbon quantum dots (CQDs), and possibly graphene quantum dots (GQDs) as observed by HR-TEM. We also confirmed that the particles are still present in significant concentrations after the rinsing process, as the signal in UV-Visible and Photoluminescence are very clear. And at last, the contribution of the solution pH is clear, the emission wavelength suffers a redshift, and increases the range of excitation wavelengths to higher energy wavelenths, probably due to a protonation occurring for lower pH. CQDs and GQDs are very promissing in biosensors applications, the low specific capacitance may enable an application in electrochemical biosensor, and the fluorescence caused by surface functional groups for fluorescent nanosensors. These potential applications will be further comprehended in next studies.

## 5. ACKNOWLEDGMENTS

The authors thanks for the financial support to “National Council for Scientific and Technological Development – CNPq, Process: 302584/2023-4, 404412/2024-6 and Universidade Federal do ABC. Finally, to CEM at the Universidade Federal do ABC for the equipments used for samples characterizations.